\input harvmac.tex

\def\np#1#2#3{Nucl. Phys. {\bf B{#1}} ({#2}) {#3}}
\def\pl#1#2#3{Phys. Lett. {\bf B{#1}} ({#2}) {#3}}
\def\prd#1#2#3{Phys. Rev. {\bf D{#1}} ({#2}) {#3}}

\def\one{\bf{1}}
\def\two{\bf{2}}
\def\three{\bf{3}}
\def\four{\bf{4}}
\def\five{\bf{5}}
\def\six{\bf{6}}
\def\seven{\bf{7}}

\Title{\vbox{\baselineskip12pt\hbox{hep-th/9804060}
\hbox{RU-98-12}}}
{\vbox{
\centerline{Spectrum of Chiral Operators in}
\vskip 10pt
\centerline{Strongly Coupled Gauge Theories}
}}
\centerline{Rami Entin and Jaume Gomis}
\vskip 15pt
\centerline{\it Department of Physics and Astronomy}
\centerline{\it Rutgers University }
\centerline{\it Piscataway, NJ 08855-0849}
\centerline{and}
\centerline{\it Institute for Theoretical Physics}
\centerline{\it University of California, Santa Barbara, CA 93106}

\medskip
\centerline{\tt rami, jaume@physics.rutgers.edu}
\medskip
\bigskip
\noindent

We analyze the large $N$ spectrum of chiral primary operators of three
dimensional   
fixed points of the renormalization group. Using the space-time picture of the
fixed points
and the correspondence between anti-de Sitter compactifications and 
conformal field theories  we are able to extract the dimensions of operators
in short superconformal multiplets. We write down some of these operators in 
terms of short distance theories  flowing to these non-trivial fixed
points in the   
infrared.

\bigskip
 
\Date{April 1998}

\newsec{Introduction}

Recently, new ways of analyzing conformal field theories have been
proposed
\nref\juan{J.M. Maldacena, ``The Large N Limit of Superconformal Field Theories and Supergravity'', hep-th/9711200.}%
\nref\FF{S. Ferrara and C. Fr\o ndsal, ``Conformal Maxwell Theory as a Singleton Field Theory on $AdS_5$, IIB Three-Branes and Duality'', hep-th/9712239.}%
\nref\GKP{S.S. Gubser, I.R. Klebanov and A.M. Polyakov, ``Gauge Theory Correlators from Noncritical String Theory'', 
hep-th/9802109.}%
\nref\WITone{E. Witten, ``Anti-de Sitter Space and Holography'', hep-th/9802150.}%
\refs{\juan,\FF,\GKP,\WITone}. Certain conformal field theories
are conjectured to be dual to certain space-time compactifications of
string theory or M-theory containing anti-de Sitter geometry in the
background \juan. Evidence for the conjecture follows from identifying
the space-time symmetries with symmetries of the conformal field theory
\nref\CRT{P. Claus, R. Kallosh and A. van Proeyen, ``M Five-Brane and Superconformal $(0,2)$ Tensor Multiplet in Six Dimensions'', hep-th/9711161,\parskip=0pt
\item{}R. Kallosh, J. Kumar and A. Rajaraman, ``Special Conformal Symmetry of World Volume Actions'', hep-th/9712073,\parskip=0pt
\item{}P. Claus, R. Kallosh, J. Kumar, P.K. Townsend and A. van Proeyen, ``Conformal Theory of M2, D3, M5 and D1+D5-Branes'', 
hep-th/9801206.}%
\nref\micha{M. Berkooz, ``A Supergravity Dual of a $(1,0)$ Field Theory in Six Dimensions'', hep-th/9802195.}%
\nref\jaume{J. Gomis, ``Anti-de Sitter Geometry and Strongly Coupled Gauge Theories'', hep-th/9803119.}
\refs{\juan,\CRT,\FF,\micha,\jaume}. Moreover, in \refs{\GKP,\WITone}
a 
precise recipe was given for computing arbitrary 
correlation functions
 in the conformal field theory in
terms of the dual space-time theory. The correlation
function of operators in the conformal field theory can be computed
by analyzing the supergravity action dependence on the boundary of the anti-de
 Sitter geometry. The correspondence between space-time states and
operators in 
the conformal field theory can be used to test or predict results in
 conformal field theories. D-brane
technology and string duality have recently been used to construct
conformal field 
theories and their conjectured space-time duals
\nref\KS{S. Kachru and E. Silverstein, ``4D Conformal Field Theories and Strings on Orbifolds'', hep-th/9802183.}%
\nref\LNV{A. Lawrence, N. Nekrasov and C. Vafa, ``On Conformal Field Theories in Four Dimensions'', hep-th/9803015.}%
\nref\FZ{S. Ferrara and A. Zaffaroni, ``$N=1,2\, 4D$ Superconformal Field Theories and Supergravity in $AdS_5$'', hep-th/9803060.}%
\nref\BKV{M. Bershadsky, Z. Kakushadze and C. Vafa, ``String Expansions as Large N Expansions of Gauge Theories'', hep-th/9803076.}%
\nref\AOY{O. Aharony, Y. Oz and Z. Yin, ``M Theory on $AdS_p\times S^{11-p}$ and Superconformal Field Theories'', hep-th/9803051.}%
\nref\MINone{S. Minwalla, ``Particles on $AdS_{4/7}$ and Primary Operators on $M_{2/5}$ Branes Worldvolumes'', hep-th/9803053.}%
\nref\LR{R.G. Leigh and M. Rozali, ``The Large N Limit of the $(2,0)$ Superconformal Field Theory'', hep-th/9803068.}%
\nref\HALone{E. Halyo, ``Supergravity on $AdS_{4/7}\times S^{7/4}$ and M branes'', hep-th/9803077.}%
\nref\FKPZ{S. Ferrara, A. Kehagias, H. Partouche and A. Zaffaroni, ``Membranes and Five-Branes with Lower Supersymmetry and Their $AdS$ Supergravity Duals'', 
hep-th/9803109.}%
\nref\OT{Y. Oz and J. Terning, ``Orbifolds of $AdS_5\times S^5$ and $4d$ Conformal Field Theories'', hep-th/9803171.}%
\nref\HALtwo{E. Halyo, ``Supergravity on $AdS_{5/4}\times$ Hopf Fibrations and Conformal Field Theories'', hep-th/9803193.}%
\nref\kaku{Z. Kakushadze, ``Gauge Theories from Orientifolds and Large N Limit'', hep-th/9803214.}%
\nref\ferrkeparza{S. Ferrara, A. Kehagias, H. Partouche and A. Zaffaroni, 
hep-th/9804006.}%
$[8-20,7]$. Related work which has appeared recently is
\nref\SS{K. Sfetsos and K. Skenderis, ``Microscopic Derivation of the Beckenstein-Hawking Entropy Formula for Nonextremal Black Holes'', hep-th/9711138.}%
\nref\FFtwo{S. Ferrara and C. Fr\o ndsal, ``Gauge Fields as Composite Boundary Excitations'', hep-th/9802126.}%
\nref\BPSone{H.J. Boonstra, B. Peeters and K. Skenderis, ``Branes and Anti-de Sitter Space-Times'', hep-th/9801076.}%
\nref\IMSS{N. Itzhaki, J.M. Maldacena, J. Sonnenschein and S. Yankielowicz, ``Supergravity and the Large N Limit of Theories with Sixteen Supercharges'', hep-th/9802042.}%
\nref\GM{M. G\"unaydin and D. Minic, ``Singletons, Doubletons and M Theory'',  hep-th/9802047.}%
\nref\HO{G.T. Horowitz and H. Ooguri, ``Spectrum of Large N Gauge Theory from Supergravity'', hep-th/9802116.}%
\nref\BL{V. Balasubramanian and F. Larsen, ``Near Horizon Geometry and Black Holes in Four Dimensions'', hep-th/9802198.}%
\nref\FFZ{S. Ferrara, C. Fr\o ndsal and A. Zaffaroni, ``On $N=8$ Supergravity on $AdS_5$ and $N=4$ Superconformal Yang-Mills Theory'', hep-th/9802203.}%
\nref\RY{S.-J. Rey and J. Yee, ``Macroscopic Strings as Heavy Quarks in Large N gauge Theory and Anti-de Sitter Supergravity'', hep-th/9803001.}%
\nref\mal{J. Maldacena, ``Wilson Loops in Large N Gauge Theories'', hep-th/9803002.}%
\nref\flato{M. Flato and C. Fr\o ndsal, ``Interacting Singletons'', hep-th/9803013.}%
\nref\GHKK{S.S. Gubser, A. Hashimoto, I.R. Klebanov and M. Krasnitz, ``Scalar Absorption and the Breaking of the World Volume Conformal Invariance'', 
hep-th/9803023.}%
\nref\AV{I.Ya. Aref'eva and I.V. Volovich, ``On Large N Conformal Field Theories in Anti-de Sitter Space and Singletons'', hep-th/9803028.}%
\nref\CCD{L. Castellani, A. Ceresole, R. D'Auria, S. Ferrara, P. Fr\'e and
M. Trigiante, ``$G/H$ M-branes and $AdS_{p+2}$ Geometries'',
hep-th/9803039.}%
\nref\DLP{M.J. Duff, H. L\"u and C.N. Pope, ``$AdS_5\times S^5$ Untwisted'', hep-th/9803061.}%
\nref\arvind{A. Rajaraman, ``Two Form Fields and the Gauge Theory Description of Black Holes'', hep-th/9803082.}%
\nref\HR{G.T. Horowitz and S.F. Ross, ``Possible Resolution of Black Holes Singularities from Large N Gauge Theory'', hep-th/9803085.}%
\nref\BB{E. Bergshoeff and K. Behrndt, ``D-Instantons and Asymptotic Geometries'', hep-th/9803090.}%
\nref\mina{J.A. Minahan, ``Quark-Monopoles Potentials in Large N Super Yang-Mills'', hep-th/9803111.}%
\nref\WITtwo{E. Witten, ``Anti-de Sitter Space, Thermal Phase Transition and Confinement in Gauge Theories'', hep-th/9803131.}%
\nref\RTY{S.-J. Rey, S. Theisen and J.T. Yee, ``Wilson-Polyakov Loop at Finite Temperature in Large N Gauge Theory and Anti-de Sitter Supergravity'', hep-th/9803135.}%
\nref\BISS{A. Brandhuber, N. Itzhaki, J. Sonnenschein and S. Yankielowicz, ``Wilson Loops in the Large N Limit at Finite Temperature'', hep-th/9803137.}%
\nref\guna{M. G\"unaydin, ``Unitary Supermultiplets of $Osp(1/32,R)$ and M-Theory'', hep-th/9803167.}%
\nref\AF{L. Andrianopoli and S. Ferrara, ``K-K Excitations on $AdS_5\times S^5$ as $N=4$ `Primary' Superfields'', hep-th/9803171.}%
\nref\BGS{B. Brandt, J. Gomis and J. Simon, ``D Strings on Near Horizon Geometries and Infinite Conformal Symmetry'', hep-th/9803196.}%
\nref\VOL{A. Volovich, ``Near Anti-de Sitter Geometry and Corrections to the Large N Wilson Loop'', hep-th/9803220.}%
\nref\BPStwo{H.J. Boonstra, B. Peeters and K. Skenderis, ``Brane Intersections, Anti-de Sitter Space-Times and Dual Superconformal Theories'', hep-th/9803231.}%
\nref\HSeft{M. Henningson and K. Sfetsos, ``Spinors and the 
ADS/CFT Correspondence'', hep-th/9803251.}%
\nref\mli{M. Li, `` 't Hooft Vortices on D-Branes'', hep-th/9803252.}%
\nref\confino{A. Brandhuber, N. Itzhaki, J. Sonnenschein and 
S. Yankielowicz, 
``Wilson Loops, Confinement 
and Phase Transitions in Large N Gauge Theories from 
Supergravity'', hep-th/9803263.}%
\nref\emparan{R. Emparan, ``AdS Membranes Wrapped on Surfaces of Arbritary 
Genus'', hep-th/9804031.}%
$[21-51]$.

In \refs{\FKPZ,\jaume}, M-theory duals to three dimensional fixed
points
\nref\SEIBone{N. Seiberg, ``IR Dynamics on Branes and Space-Time Geometry'', \pl{384}{1996}{81},
hep-th/9606017.}%
\nref\IS{K. Intriligator and N. Seiberg, ``Mirror Symmetry in Three-Dimensional Gauge Theories'', \pl{387}{1996}{513}, hep-th/9607207}
\nref\BHO{J. de Boer, K. Hori, H. Ooguri, Y. Oz and Z. Yin, ``Mirror Symmetry in Three-Dimensional Field Theories, $SL(2,Z)$ and D-Brane Moduli Spaces'', 
\np{493}{1997}{148}, hep-th/9612131.}%
\refs{\SEIBone,\IS,\BHO} of the renormalization group were given. These
theories can be realized as world-volume theories of M2-branes sitting
at an ADE singularity. These strongly coupled gauge theories are
difficult to analyze and one might hope to learn more about them by
using the AdS-CFT correspondence. 

In this paper, we identify part of the spectrum of primary chiral
operators of three dimensional superconformal field theories. This is
done by using the M-theory description of the theory at the fixed
points
\refs{\FKPZ,\jaume} and identifying the space-time fields in short
supersymmetry  multiplets. These fields are in one to one
correspondence with chiral operators of the superconformal algebra.

In section two we briefly review the Kaluza Klein spectrum of eleven
dimensional supergravity on $AdS_4\times S^7$
\nref\fluct{B. Biran, A. Casher, F. Englert, M. Rooman and P. Spindel,
``The Fluctuating Seven Sphere in Eleven Dimensional Supergravity'',
\pl{134}{1984}{179}.}%
\nref\spectb{L. Castellani, R. D'Auria, P. Fr\'e, K. Pilch and P. van Nieuwenhuizen,
``The Bosonic Mass Formula for Freund-Rubin Solutions of $d=11$
Supergravity General Coset Manifolds'', Class. Quant. Grav. {\one}
(1984) 339.}%
\refs{\fluct,\spectb}. In section three the space-time  duals
\refs{\FKPZ,\jaume} 
of the non-trivial fixed points of the renormalization group are
introduced. In this section we analyze the  states that survive the
orbifold projection and identify the subset of states with the right
quantum numbers to be in reduced supersymmetry multiplets. In section
four dual  short distance theories, related by mirror symmetry
\refs{\SEIBone,\IS,\BHO},  which flow to these fixed points in the
infrared are introduced. We give candidate chiral operators in the
superconformal field theory using fields in the short distance
theories. Section five  contains conclusions. 

\newsec{Kaluza-Klein Spectrum of Supergravity on $AdS_4\times S^7$}

Kaluza-Klein harmonics in supergravity play a pivotal role in
establishing the AdS-CFT correspondence \WITone.  They belong to small
representations of the supersymmetry algebra in the maximally
supersymmetric case
\nref\FN{D.Z. Freedman and A. Nicolai, ``Multiplet Shortening in $Osp(N,4)$'', \np{237}{1984}{342}.}%
\nref\heiden{W. Heidenreich, ``All Linear Unitary Irreducible representations of de Sitter Supersymmetry with Positive Energy'', \pl{110}{1982}{461}.}%
\refs{\FN,\heiden}. As such, many of their
properties are protected from quantum and stringy corrections. In
particular their masses are given by the eigenvalues of the relevant
Laplace operator. Therefore, operators in the superconformal field
theory which couple to them will be chiral since their dimensions will
be protected from quantum corrections \WITone. Chiral operators are in
small representations of the superconformal algebra (see 
\nref\SEIBtwo{N. Seiberg, ``Notes on Theories with Sixteen Supercharges'', hep-th/9705117.}%
\nref\MINtwo{S. Minwalla, ``Restrictions Imposed by Superconformal Invariance on Quantum Field Theories'', hep-th/9712074.}%
\refs{\SEIBtwo,\MINtwo} for details). Their dimensions
saturate an inequality of the form
\eqn\ineq{D({\cal{O}})\geq f(\{L_i\},\{R_i\}),}
where $\{L_i\},\{R_i\}$ are the Lorentz and R-symmetry quantum
numbers. The function $f$ is determined by requiring the
representation of the superconformal algebra carried by these
operators to be unitary.

One can exploit the AdS-CFT correspondence to verify or predict the
spectrum of chiral operators of a given superconformal field theory
\refs{\WITone,\AOY,\MINone,\LR,\HALone,\OT}. In this section we
will review the Kaluza-Klein spectrum of supergravity in $AdS_4\times
S^7$ \refs{\fluct,\spectb} which we will use in the following sections
to determine the spectrum of chiral operators of three dimensional
strongly coupled gauge theories. In
\refs{\AOY,\HALone} this spectrum was 
used to determine the chiral operators of an ${\cal N}=8$ three
dimensional superconformal field theory.

The spectrum consists of three families of scalar excitations, two
families of pseudoscalar excitations, two families of vector
excitations, one family of axial vectors
 and one family of spin 2 excitations.  A given Kaluza
Klein excitation transforms in a particular representation of the
$SO(8)$ isometry group of $S^7$ and carries the Lorentz quantum
numbers of the family to which it belongs.  The mass\foot{The mass
squared of a field is the eigenvalue of the relevant differential
operator.} of a given space-time field determines the renormalization
group flow properties of the corresponding operator in the conformal
field theory \WITone. The dimension  of the operator that couples to a
space-time $p$-form is given by $(\Delta +p) (\Delta +p  - d)=m^2$,
where $d$ is  the number of dimensions in which the conformal field
theory lives.  Thus tachyonic, massless and massive states couple to
relevant, marginal and irrelevant operators respectively \WITone. In
what follows, we shall only consider those families which contain
tachyonic and massless modes corresponding to relevant and marginal
operators. These are listed below: 

\noindent
$\bullet$ Scalar:

\noindent
$m^2={1\over 4}k(k-6),\ k\geq 2.$ These states transform in the $k$-th
symmetric traceless representation of $SO(8)$; their Dynkin labels are
$(k,0,0,0)$. The dimensions of the corresponding chiral operators in
the CFT are $\Delta={k\over 2}$. 

\noindent
$\bullet$ Pseudo-scalar:

\noindent
$m^2={1\over 4}((k-1)(k+1)-8), \ k\geq 1.$ These states transform in
the product representation of the ${\bf 35_c}$ with the $k-1$-th
symmetric traceless representation of $SO(8)$; their Dynkin labels are
$(k-1,0,2,0)$. The dimensions of the corresponding chiral operators in
the CFT are $\Delta={k+3\over 2}$. 

\noindent
$\bullet$ Vector

\noindent
$m^2={1\over 4}(k^2-1),\ k\geq 1.$ These states transform in the
product representation of the ${\bf 28}$ with the $k-1$-th symmetric
traceless representation of $SO(8)$; their Dynkin labels are
$(k-1,1,0,0)$. The dimensions of the corresponding chiral operators in
the CFT are $\Delta={k+3\over 2}$. 

\noindent
$\bullet$ Graviton

\noindent
$m^2={1\over 4}k(k+6),\ k\geq 0.$ These states transform in the $k$-th
symmetric traceless representation of $SO(8)$; their Dynkin labels are
$(k,0,0,0)$. The dimensions of the corresponding chiral operators in
the CFT are $\Delta={k+6\over 2}$.

States in different Kaluza Klein towers can be combined in
supermultiplets by the action of the supersymmetry generators. The
operators in the first scalar family are the only ${\cal N}=8$
superconformal primaries 
\nref\GW{M. G\"unaydin and N.P. Warner, ``Unitary Supermultiplets of OSp(8/4,R) and the Spectrum of the $S^7$ Compactification of 11-Dimensional Supergravity'', \np{272}{1986}{99}.}%
\GW.

\newsec{States in reduced supersymmetry multiplets}

In \refs{\FKPZ,\jaume} supergravity duals of ${\cal N}=4$ three
dimensional fixed point theories with ADE global symmetries were
given. These theories have dual short distance descriptions related by
three dimensional mirror symmetry \refs{\SEIBone,\IS,\BHO}. They can
be realized as world volume theories of brane configurations in
space-time.  Taking the limit as in \juan\ of the supergravity
description yields the  space-time theory dual to the fixed point. The
space-time  compactification can be  read from the near horizon
geometry of the brane configuration.

The brane configuration corresponding to the fixed point is the theory
on $N$ M2-branes at a $C^2/\Gamma$ singularity\foot{The discrete group
$\Gamma\in SU(2)$ is one of the ADE discrete groups.}.  The near
horizon geometry of this  brane configuration  is $AdS_4\times
S^3\times_fD_4/\Gamma$, where the $S^3$ is fibered over the $\Gamma$
quotiented four-disk. Therefore, the theory dual to the fixed point is
M-theory on $AdS_4\times S^3\times_fD_4/\Gamma$
\jaume. We would like to emphasize the need to include all the degrees
of freedom of M-theory and not just the eleven dimensional
supergravity ones in order to regulate the singular geometry at the
origin of the disk. In \jaume\ all the symmetries of the fixed point
were identified  with symmetries in space-time, including the ADE
global symmetries.

The procedure for finding the chiral spectrum is analogous to that of
\OT. We know that the space-time realization of the fixed point is the
theory of $N$ M2-branes sitting at an ADE singularity. The presence of
the ADE singularity breaks the transverse $SO(8)$ global symmetry to
$SO(4)_{ADE}\times SO(4)_{E}$. Furthermore, the discrete group
$\Gamma$ acts in $SU(2)_{\Gamma} \subset SO(4)_{ADE}$, which breaks
the $SO(8)$ isometry group of $S^7$ to $SU(2)_{ADE}\times
SU(2)_{L}\times SU(2)_R$. The unbroken supersymmetry charges of the
theory on the M2-branes transform in a $({\bf 2}, {\bf 1}, {\bf 2})$
representation of $SU(2)_{ADE}\times SU(2)_{L}\times
SU(2)_R$. Therefore, we identify $SU(2)_{ADE}\times SU(2)_{R}$ as the
R-symmetry of the three dimensional fixed point theory. The $SU(2)_L$
symmetry acts as an additional global symmetry group and will be kept
for most of the analysis. 

The geometry of the configuration above is singular, since the action
of $\Gamma$ leaves an invariant $S^3$ at the origin. However, we assume 
that the Kaluza-Klein modes invariant under this action
will correspond to chiral operators in the field theory.  Therefore,
one must decompose the Kaluza Klein $SO(8)$ representations of
$AdS_4\times S^7$ under $SU(2)_{\Gamma}\times SU(2)_{ADE}\times
SU(2)_{L}\times SU(2)_R$, and keep only those states in
$SU(2)_{\Gamma}$ representations which have neutral components under
the  action of $\Gamma$ . As will be elaborated later on, only a
subset of the states surviving the projection will have the right
quantum numbers to be chiral operators of the superconformal field
theory.
 
The branching rules, $SO(8)\rightarrow SU(2)_{\Gamma}\times
SU(2)_{ADE}\times SU(2)_{L}\times SU(2)_R$, for the $SO(8)$
representations corresponding to relevant and marginal operators are
given in the appendix\foot{See 
\nref\patera{W.G. McKay and J. Patera, ``Tables of Dimensions, Indices and
Branching Rules for  Representations of Simple Lie Algebras'', 
Marcell Dekker 1981.}%
\patera.}. Let us consider the case $\Gamma=A_{k-1}=Z_{k}$
which acts in the  $\two$ of $SU(2)_{\Gamma}$ as 
\eqn\action{\eqalign{
Z^1\rightarrow &\alpha Z^1\cr Z^2\rightarrow &\alpha^{-1}Z^2,}}  where
$\alpha=e^{2\pi i \over k}$. The $SU(2)_{\Gamma}$ representations
which have  an invariant component under the $Z_{k}$ action\foot{Performing the
same analysis for D and E discrete groups will just further restrict the 
allowed representations.}
are those
which are a product of an even number of $\two$'s. Therefore we will
keep only the odd  dimensional representations of $SU(2)_\Gamma$ that
appear in the branching  rules. 

The first family of scalar excitations in the previous section has
five Kaluza-Klein modes which are either tachyonic or massless. The
states that survive the projection are\foot{The triplet $({\bf a},{\bf
b},{\bf c})$ corresponds to an $SU(2)_{ADE}\times SU(2)_{L}\times
SU(2)_R$ representation.}

\noindent
$\bullet\ k=2\quad m^2=-2$

\noindent
${(\three,\one,\one)}\oplus{(\one,\three,\three)}\oplus{(\one,\one,\one)}$

\noindent
These states couple to a $\Delta=1$ operator of the fixed point
theory.

\noindent
$\bullet\ k=3\quad m^2=-{9\over{4}}$

\noindent
${(\three,\two,\two)}\oplus {(\one,\four,\four)}\oplus
{(\one,\two,\two)}$

\noindent
These states couple to a $\Delta={3\over{2}}$ operator of the fixed
point theory.
 
\noindent
$\bullet\ k=4\quad m^2=-2$

\noindent
${(\five,\one,\one)}\oplus {(\three,\three,\three)} \oplus
{(\one,\five,\five)} \oplus {(\three,\one,\one)} \oplus
{(\one,\three,\three)} 
\oplus {(\one,\one,\one)}$

\noindent
These states couple to a $\Delta=2$ operator of the fixed point
theory.

\noindent
$\bullet\ k=5\quad m^2=-{5\over{4}}$

\noindent
${(\five,\two,\two)}\oplus {(\three,\four,\four)} \oplus
{(\one,\six,\six)} \oplus {(\three,\two,\two)} \oplus
{(\one,\four,\four)}\oplus {(\one,\two,\two)}$

\noindent
These states couple to a $\Delta={5\over{2}}$ operator of the fixed
point theory. 

\noindent
$\bullet\ k=6\quad m^2=0$

\noindent
${(\seven,\one,\one)}\oplus {(\five,\three,\three)} \oplus
{(\three,\five,\five)} \oplus {(\one,\seven,\seven)} \oplus
{(\five,\one,\one)}\oplus {(\three,\three,\three)} \oplus
{(\one,\five,\five)} \oplus {(\three,\one,\one)} \oplus
{(\one,\three,\three)}
\oplus {(\one,\one,\one)}$

\noindent
These states couple to a $\Delta=3$ operator of the fixed point
theory. 

The second family of pseudoscalar excitations has three Kaluza-Klein
modes which are tachyonic or massless. The states that survive the
projection are

\noindent
$\bullet\ k=1\quad m^2=-2$

\noindent
${(\three,\three,\one)}\oplus {(\one,\one,\three)}\oplus
{(\one,\one,\one)}$

\noindent
These states couple to a $\Delta=2$ operator of the fixed point
theory. 

\noindent
$\bullet\ k=2\quad m^2=-{5\over{4}}$

\noindent
${(\three,\four,\two)}\oplus {(\three,\two,\two)}\oplus
{(\one,\two,\four)}\oplus {(\three,\two,\two)} \oplus
{(\one,\two,\two)}
\oplus {(\one,\two,\two)}$

\noindent
These states couple to a $\Delta={5\over{2}}$ operator of the fixed
point theory. 

\noindent
$\bullet\ k=3\quad m^2=0$

\noindent
${(\five,\three,\one)}\oplus {(\three,\five,\three)}\oplus
{(\three,\one,\three)}\oplus {(\three,\three,\three)} \oplus
{(\one,\three,\four)}
\oplus {(\three,\one,\three)}\oplus {(\three,\three,\three)}\oplus {(\three,\three,\one)}
\oplus{(\one,\three,\three)} \oplus {(\three,\one,\three)} \oplus {(\three,\one,\one)} \oplus {(\three,\three,\one)} \oplus {(\one,\three,\one)}
\oplus {(\one,\one,\three)} \oplus {(\one,\three,\three)} \oplus {(\one,\one,\one)}$

\noindent
These states couple to a $\Delta=3$ operator of the fixed point
theory. 

The massless vector excitation is

\noindent
$\bullet\ k=1 \quad m^2=0$

\noindent
${(\three,\one,\one)}\oplus {(\one,\one,\three)} \oplus
{(\one,\three,\one)} \oplus {(\one,\one,\one)}$

\noindent
The gauge boson couples to a $\Delta= 2$ global symmetry current
operator of the fixed point theory in the adjoint of $SO(4)$.

The massless graviton excitation is 

\noindent
$\bullet\ k=0 \quad m^2=0$

\noindent
${(\one,\one,\one)}$

\noindent
The graviton couples to the $\Delta=3$ energy-momentum tensor of the
fixed point theory.

As mentioned earlier the dimensions of superconformal chiral operators
are uniquely determined by their Lorentz and R-symmetry quantum
numbers. Therefore, only a subset of the allowed projected states will
have the correct quantum numbers to be in reduced supersymmetry
multiplets. The relation between  the dimension of chiral operators
and quantum numbers can be obtained by analyzing the allowed unitary
representations of the superconformal algebra. The theories we are
considering are three dimensional superconformal field theories with
$SO(4)$ R-symmetry. The scaling dimensions of primary chiral fields
are determined in terms of the Lorentz $Spin(3)\simeq SU(2)$ quantum
number $j$, and the $SO(4)\simeq SU(2)_{ALE}\times SU(2)_R$ highest
weights \foot{Our conventions for $h_i\ i=1,2$ are that the dimension
$d$ of an $SU(2)$ representation is $d=2h_i +1$.} $(h_1,h_2)$ as
\refs{\SEIBtwo,\MINtwo} 
\eqn\chiral{\eqalign{
D({\cal O})&=h_1+h_2\quad (j=0)\cr D({\cal O})&=1+j+h_1+h_2 \quad
(j\not= 0).}}

\noindent
Using \chiral, the superconformal primaries from the scalar family are
given by

\eqn\scalar{\vbox{
\halign{
\quad $#$\quad\hfill
&\quad $#$\quad\hfill\cr
k=2\qquad\Delta=1&({\three,\one,\one})\oplus({\one,\three,\three})\cr
k=3\qquad\Delta={3\over{2}}&({\three,\two,\two})\oplus({\one,\four,\four})\cr
k=4\qquad\Delta=2&({\three,\three,\three})\cr
k=5\qquad\Delta={5\over{2}}&({\five,\two,\two})\oplus({\three,\four,\four})\cr
k=6\qquad\Delta=3&({\five,\three,\three})\oplus({\three,\five,\five}).\cr
}}} None of the surviving states from the pseudoscalar tower have the
right  quantum numbers to be ${\cal N}=4$ superconformal
primaries. The massless vector and the graviton correspond  to
conserved currents in the superconformal field theory. They couple to
the global symmetry current in the adjoint of the R-symmetry and to
the energy momentum respectively. Since they are conserved currents,
their dimensions are  protected from quantum corrections.

\newsec{Superconformal primary operators}

In the previous section we have obtained the spectrum of states in
reduced supersymmetry multiplets. In this section we write some of the
corresponding chiral operators in terms of fields in short distance 
theories. Each fixed point has two short distance descriptions, related
by three dimensional mirror symmetry \refs{\SEIBone,\IS,\BHO}, which flow to
the fixed point in the infrared. 
 As we
shall see, some of the operators can be realized in both short
distance theories, some in only one of the mirror pairs and some of the
operators cannot be written in terms of short distance degrees of
freedom.

The short distance theories have a moduli space of vacua with a Higgs and a 
Coulomb branch which intersect at the origin. There the theories are at a
non-trivial fixed point of the renormalization group with ADE 
global symmetry. In the $Z_k$ case we are considering, the fixed point theory 
has a global $SU(k)$ symmetry. The two mirror theories are:

\noindent
$\bullet$ The theory of $N$ D2-branes at a $C^2/Z_k$
orbifold singularity. These are the quiver gauge theories considered in
\nref\kronhe{P.B. Kronheimer, ``The Construction of ALE Spaces as 
Hyperk\"a hler Quotients'', J. Diff. Geom., {\bf 29}
(1989) 665.}%
\nref\DM{M.R. Douglas and J. Moore, ``D-branes, Quivers and ALE
Instantons'', hep-th/ 9603167.}%
\nref\CVJ{C.V. Johnson and R.C. Myers, ``Aspects of Type IIB on ALE
Spaces'', \prd{55}{1997}{6382}, hep-th/9610140.}%
\refs{\kronhe,\DM,\CVJ}. 
The gauge group is $G=\left(\prod_{i=1}^kU(N)\right)/U(1)$ with $k$ 
bifundamental hypermultiplets. In ${\cal N}=1$ language, there are $2k$ chiral
superfields $Q_i,{\widetilde Q}_i^\dagger$ both transforming in the ${\bf N}_i\times{\bf{\bar{N}}}_{i+1}$ of 
$U(N)_i\times U(N)_{i+1}$.
\foot{The index $i$ is cyclic.}

\noindent
$\bullet$ The theory of $N$ D2-branes probing $k$ D6-branes in 
Type I' orientifold $T^3/Z_2\Omega$ \SEIBone. This is a $U(N)$ gauge theory
with $k$ hypermultiplets in the fundamental representation and one 
hypermultiplet in the adjoint. In ${\cal N}=1$ language, there are $2k$
chiral superfields $q_i,{\widetilde{q}}^\dagger_i$ transforming in the 
{\bf N} of $U(N)$. 

These three dimensional ${\cal N}=4$ gauge theories have an $SO(4)\simeq
SU(2)_1\times SU(2)_2$ global R-symmetry. The massless multiplets
transform in some representation of the R-symmetry which can be
inferred by  dimensional reduction of $d=6,\, {\cal N}=1$ theories. The
scalars in the vector  multiplet transform in the
$({\three,\one})$. Upon dualizing the photon along  the Coulomb branch
one gets an additional scalar transforming in the
$({\one,\one})$. Therefore along the Coulomb branch the field content
is that  of a hypermultiplet with scalars transforming in the
$({\three}\oplus{\one},{\one})$. Both scalars in the usual
hypermultiplet transform as a $({\one,\two})$. 

Mirror symmetry acts by exchanging the Coulomb (Higgs) branch of one
theory  with the Higgs (Coulomb) branch of the other. Hence, under
this symmetry the scalars in the vector (hyper) multiplet of one
theory will transform in the $SU(2)$ factor of the R-symmetry in which
the scalars in the hyper (vector) multiplets of the other theory
transform. Once the R-symmetry transformation properties of multiplets
in one theory are determined, those  of the mirror theory multiplets
are also known.

In section three we identified $SU(2)_{ADE}\times SU(2)_R$ as the R-symmetry 
of the theory at the fixed point. In the quiver construction 
\refs{\kronhe,\DM,\CVJ} the hypermultiplets are naturally charged
under the unbroken $SU(2)_{ADE}$. Therefore, the fields in the quiver
theory transform  as
\eqn\transa{\eqalign{
\Phi^A_i\oplus\Phi_i&\rightarrow({\one,\three\oplus\one})\qquad A=1,2,3\cr
Q_i,{\widetilde Q}_i&\rightarrow({\two,\one}),}}
where $\Phi^A_i$ are the vector multiplet scalars and $\Phi_i$ is the
dualized 
photon. As explained above this implies that the mirror theory fields transform
as
\eqn\transb{\eqalign{
\phi^A\oplus\phi& \rightarrow({\three\oplus\one,\one})\qquad A=1,2,3\cr
q_i,{\widetilde q}_i& \rightarrow ({\one,\two}),}}
where $\phi^A,\phi$ are the vector multiplet scalars and the dual photon.

We want to write down gauge invariant operators corresponding to the
supergravity states in small supersymmetry multiplets \scalar. A careful 
analysis shows that not all of them can be written in terms of the short 
distance fields. Concretely, while one can always write combination with the
right dimension and R-symmetry charges in at least one of the mirror theories, 
it is not always possible to do it in a gauge invariant way.

As an example, consider the $\Delta={3\over{2}}$ operator appearing in \scalar,
with R-symmetry numbers $({\one,\four})$\foot{Recall that the first and the 
third entries in each triplet are the R-symmetry quantum numbers.}. To write 
this operator in the quiver theory, one has to fully symmetrize three 
hypermultiplet scalar fields and this cannot be made in a gauge invariant way.
Obviously, this is impossible in the probe theory as well.
The same conclusion holds for all operators in the second and fourth lines of 
\scalar. 

The operators which can be written in one or both mirror pairs are given by:

\noindent
$\bullet$ Quiver theory
\eqn\opsone{\eqalign{
&\Delta=1\quad({\three,\one})\qquad
\sum_{i=1}^k\sigma^A_{\alpha\beta}\left(Q_i^\alpha{\widetilde Q}_i^\beta\right)\cr
&\Delta=2\quad({\three,\three})\qquad
\sum_{i=1}^k\sigma^A_{\alpha\beta}\left({\widetilde Q}_i^\alpha\Phi^B_{i+1}Q^\beta_i
-{\widetilde Q}_{i+1}^\alpha\Phi^B_{i+1}Q^\beta_{i+1}\right)\cr
&\Delta=3\quad({\three,\five})\qquad
\sum_{i=1}^k\sigma^A_{\alpha\beta}\left({\widetilde Q}_i^\alpha\Phi^{(B}_{i+1}\Phi^{C)}_{i+1}Q^\beta_i
-{\widetilde Q}_{i+1}^\alpha\Phi^{(B}_{i+1}\Phi^{C)}_{i+1}Q^\beta_{i+1}\right)\cr
&\Delta=3\quad({\five,\three})\qquad
\sum_{i=1}^k\sigma^{(A}_{\alpha\beta}\sigma^{C)}_{\gamma\delta}\left({\widetilde Q}_i^\alpha\Phi^B_{i+1}Q^\beta_i{\widetilde Q}_i^\gamma Q^\delta_i
-{\widetilde Q}_{i+1}^\alpha\Phi^B_{i+1}Q^\beta_{i+1}{\widetilde Q}_{i+1}^\gamma Q^\delta_{i+1}
\right)
}}

\noindent
$\bullet$ Probe theory
\eqn\opstwo{\eqalign{
&\Delta=1\quad({\one,\three})\qquad
\sum_{i=1}^k\sigma^A_{\alpha\beta}\left(q_i^\alpha{\widetilde q}^\beta_i\right)\cr
&\Delta=2\quad({\three,\three})\qquad
\sum_{i=1}^k\sigma^A_{\alpha\beta}\left({\widetilde q}_i^\alpha\phi^B q^\beta_i\right)\cr
&\Delta=3\quad({\three,\five})\qquad
\sum_{i=1}^k\sigma^A_{\alpha\beta}\left({\widetilde q}_i^\alpha\phi^{(B}\phi^{C)}q^\beta_i\right)\cr
&\Delta=3\quad({\five,\three})\qquad
\sum_{i=1}^k\sigma^{(A}_{\alpha\beta}\sigma^{C)}_{\gamma\delta}\left({\widetilde q}_i^\alpha\phi^B q^\beta_i{\widetilde q}_i^\gamma q^\delta_i
\right)
}}

In the above expressions we used the Pauli matrices $\sigma^A$ for both $SU(2)_{ADE}$ and $SU(2)_R$. The gauge and $SU(2)_L$ indices are suppressed.

\newsec{Conclusions}

Maldacena's \juan\ conjecture of the AdS-CFT correspondence opens a new 
avenue of understanding for non-trivial conformal field theories. By performing 
rather simple calculations in supergravity one can infer information about field 
theories that might not be easily obtainable otherwise.
 In this 
paper we find part of the spectrum of chiral operators of three dimensional
superconformal field theories. This is done by using the space-time picture
of the conformal field theory. Analyzing which supergravity states are in
short supersymmetry multiplets yields the spectrum of chiral operators of the
three dimensional theories. Furthermore, we are able to realize some of these
operators in terms of short distance theories which flow to these 
superconformal fixed points in the infrared. 

\vfill\eject

\centerline{\bf Acknowledgments}

We would like to thank O. Aharony, T. Banks, J. Maldacena,  Y. Oz, S. Shenker 
and the other participants of the ``Duality in String Theory''
workshop for
discussions. We are grateful to everyone in  the Institute for 
Theoretical Physics for
hospitality during this work. This research was supported in part by the
National Science Foundation under Grant No. PHY94-07194.

\vfill
\eject

\appendix{A}{$SO(8)$ branching rules}

$$\vbox{
\offinterlineskip
\halign{
\strut
\vrule\quad $#$\quad\hfill
&\vrule\quad $#$\quad\hfill
&\vrule\quad $#$\quad\hfill\vrule\cr
\noalign{\hrule}
&SO(8)&SU(2)^4 \cr
\noalign{\hrule}
{\rm{scalar}}&&\cr
\noalign{\hrule}
m^2=-2&\eqalign{&(2,0,0,0)\cr&\bf{35_v}}&\eqalign{
&(\three,\three,\one,\one)\oplus(\two,\two,\two,\two)\oplus(\one,\one,\three,\three)\oplus(\one,\one,\one,\one)
}\cr 
\noalign{\hrule}
m^2=-{9\over{4}}&\eqalign{&(3,0,0,0)\cr &\bf{112_v}}&\eqalign{
&(\four,\four,\one,\one)\oplus(\three,\three,\two,\two)\oplus(\two,\two,\three,\three)\oplus(\one,\one,\four,\four)\oplus\cr
&(\two,\two,\one,\one)\oplus(\one,\one,\two,\two)
}\cr
\noalign{\hrule}
m^2=-2&\eqalign{&(4,0,0,0)\cr&\bf{294_v}}&\eqalign{
&(\five,\five,\one,\one)\oplus(\four,\four,\two,\two)\oplus(\three,\three,\three,\three)\oplus(\two,\two,\four,\four)\oplus\cr
&(\one,\one,\five,\five)\oplus(\three,\three,\one,\one)\oplus(\two,\two,\two,\two)\oplus(\one,\one,\three,\three)\oplus\cr&
(\one,\one,\one,\one)
}\cr
\noalign{\hrule}
m^2=-{5\over{4}}&\eqalign{&(5,0,0,0)\cr&\bf{672^\prime_v}}&\eqalign{
&(\six,\six,\one,\one)\oplus(\five,\five,\two,\two)\oplus(\four,\four,\three,\three)\oplus(\three,\three,\four,\four)\oplus\cr
&(\two,\two,\five,\five)\oplus(\one,\one,\six,\six)\oplus(\four,\four,\one,\one)\oplus(\three,\three,\two,\two)\oplus\cr
&(\two,\two,\three,\three)\oplus(\one,\one,\four,\four)\oplus(\two,\two,\one,\one)\oplus(\one,\one,\two,\two)
}\cr
\noalign{\hrule}
m^2=0&\eqalign{&(6,0,0,0)\cr&\bf{1386_v}}&\eqalign{
&(\seven,\seven,\one,\one)\oplus(\six,\six,\two,\two)\oplus(\five,\five,\three,\three)\oplus(\four,\four,\four,\four)\oplus\cr
&(\three,\three,\five,\five)\oplus(\two,\two,\six,\six)\oplus(\one,\one,\seven,\seven)\oplus(\five,\five,\one,\one)\oplus\cr
&(\four,\four,\two,\two)\oplus(\three,\three,\three,\three)\oplus(\two,\two,\four,\four)\oplus(\one,\one,\five,\five)\oplus\cr
&(\three,\three,\one,\one)\oplus(\two,\two,\two,\two)\oplus(\one,\one,\three,\three)\oplus(\one,\one,\one,\one)
}\cr
\noalign{\hrule}
{\rm{pseudoscalar}}&&\cr
\noalign{\hrule}
m^2=-2&\eqalign{&(0,0,2,0)\cr&\bf{35_c}}&\eqalign{
&(\one,\three,\three,\one)\oplus(\two,\two,\two,\two)\oplus(\three,\one,\one,\three)\oplus(\one,\one,\one,\one)
}\cr
\noalign{\hrule}
m^2=-{5\over{4}}&\eqalign{&(1,0,2,0)\cr&\bf{224_{cv}}}&\eqalign{
&(\two,\four,\three,\one)\oplus(\one,\three,\four,\two)\oplus(\three,\three,\two,\two)\oplus(\four,\two,\one,\three)\oplus\cr
&(\two,\two,\three,\three)\oplus(\three,\one,\four,\two)\oplus(\one,\three,\two,\two)\oplus(\two,\two,\one,\three)\oplus\cr
&(\two,\two,\three,\one)\oplus(\three,\one,\two,\two)\oplus(\two,\two,\one,\one)\oplus(\one,\one,\two\,\two)
}\cr
\noalign{\hrule}
m^2=0&\eqalign{&(2,0,2,0)\cr&\bf{840^\prime_s}}&\eqalign{
&(\three,\five,\three,\one)\oplus(\two,\four,\four,\two)\oplus(\four,\four,\two,\two)\oplus(\one,\three,\five,\three)\oplus\cr
&(\five,\three,\one,\three)\oplus(\three,\three,\three,\three)\oplus(\four,\two,\two,\four)\oplus(\two,\two,\four,\four)\oplus\cr
&(\three,\one,\three,\five)\oplus(\two,\four,\two,\two)\oplus(\three,\three,\one,\three)\oplus(\one,\three,\three,\three)\oplus\cr
&(\three,\three,\three,\one)\oplus(\two,\two,\two,\four)\oplus(\four,\two,\two,\two)\oplus(\two,\two,\four,\two)\oplus\cr
&(\three,\one,\three,\three)\oplus(\one,\three,\one,\three)\oplus(\three,\three,\one,\one)\oplus(\one,\three,\three,\one)\oplus\cr
&(\two,\two,\two,\two)\oplus(\two,\two,\two,\two)\oplus(\three,\one,\three,\one)\oplus(\three,\one,\one,\three)\oplus\cr
&(\one,\one,\three,\three)\oplus(\one,\one,\one,\one)
}\cr
\noalign{\hrule}
{\rm{vector}}&&\cr
\noalign{\hrule}
m^2=0&\eqalign{&(0,1,0,0)\cr&\bf{28}}&\eqalign{
&(\two,\two,\two,\two)\oplus(\one,\three,\one,\one)\oplus(\one,\one,\one,\three)\oplus(\one,\one,\three,\one)\oplus\cr
&(\three,\one,\one,\one)
}\cr
\noalign{\hrule}
}}$$

\listrefs

\end